# Digital Neuron: A Hardware Inference Accelerator for Convolutional Deep Neural Networks

Hyunbin Park, Dohyun Kim, and Shiho Kim

*Abstract*— We propose a Digital Neuron, a hardware inference accelerator for convolutional deep neural networks with integer inputs and integer weights for embedded systems. The main idea behind this concept is to reduce the circuit area and power consumption by manipulating the dot products between the input feature and weight vectors using barrel shifters and parallel adders. The reduced area allows more computational engines to be mounted on an inference accelerator, resulting in a high throughput compared to that of other hardware accelerators. We transformed the multiplication of two 8-bit integers into 3-partial sub-integers and verified it does not cause a significant loss in the inference accuracy compared to that in the 32-bit floating point calculation. The proposed digital neuron can perform 800 Multiply-and-Accumulation (MAC) operations in one clock for convolution as well as full-connection. This paper provides a scheme that reuses the input, weight, and output of all layers to reduce DRAM access. In addition, it proposes a configurable architecture that can enable the inference of adaptable feature of convolutional neural networks. The throughput of the digital neuron in terms of watts is 754.7 GMACs/W.

## I. Introduction

DEEP learning has revolutionized the computer industry over the last decades, pushing the accuracy of image recognition beyond that of humans [1]. However, the computational effort of current neural networks depends on the power-hungry parallel floating-point processors or General-Purpose computing on Graphics Processing Units (GP-GPUs). Recent developments in hardware accelerators have significantly reduced energy consumption; however, devices with embedded systems such as drones, smart robots, and autonomous vehicles are still facing hard limitations caused by power budget for ultralow power applications.

Several studies have adopted a strategy for training using a desktop or server and inference using a hardware accelerator in embedded devices [2]-[4]. To utilize the hardware inference accelerator in an embedded system, the neural weights in the single- or half-precision Floating Point format (FP32 or FP16) should be converted into integers.

The key issues with hardware accelerators in embedded systems are accuracy loss, power consumption, circuit area, and throughput. To improve the throughput and power consumption, several studies [5]-[8] have reduced the bit-width of the input and weight compared to those in the conventional FP32 format. The reduction in the bit-width decreases circuit area of computational engine, which therefore allows more computational engines to be mounted in a limited area. Therefore, it can increase the throughput. Gysel [5] suggested quantizing inputs and weights into 8-bits in forward propagation, resulting in accuracy loss of less than 1 % in LeNet-5 [9], AlexNet [10], and SqueezeNet [11].

TABLE I
Simulated inference accuracy of MNIST handwritten number in LeNet-5 [9] according to resolution of weights, where pre-trained weights in the network with FP32 are quantized to integer weights.

|  | Inference accuracy |
|---|---|
| 32-bit floating-point representation | 99.10 % |
| 8-bit integer weights | 99.10 % |
| 5-bit integer weights | 98.95 % |
| Proposed 5-bit integer weights with two partial sub-integers | 98.92 % |
| Proposed 8-bit integer weights with three partial sub-integers | 99.10 % |

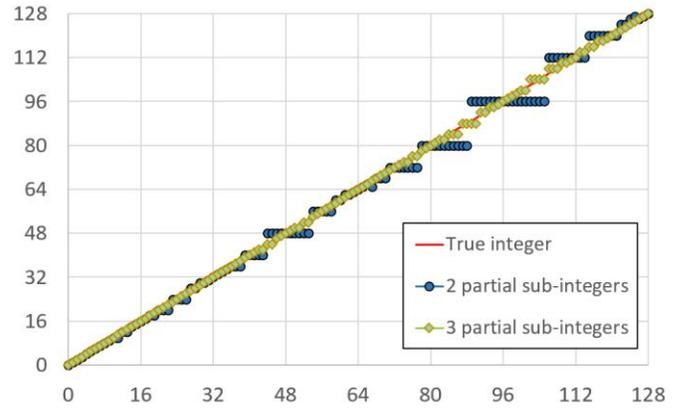

Fig. 1 Representation of integers from 0 to 128 partitioned into 2 or 3 partial sub-integers.

"This research was supported by the MSIT(Ministry of Science and ICT), Korea, under the ICT Consilience Creative program(IITP-2018-2017-0-01015) supervised by the IITP(Institute for Information & communications Technology Promotion), by Institute for Information & communications Technology Promotion(IITP) grant funded by the Korea government(MSIP) (No.2017-0-00244, HMD Facial Expression Recognition Sensor and Cyber-interaction Interface Technology)

The authors are with the School of Integrated Technology, Yonsei University, Seoul, Korea (e-mail: bin9000@yonsei.ac.kr; kimdh5032@naver.com; shiho@yonsei.ac.kr).

BinaryConnect (BC) [12] reduced the bit-width of weights even further to 1-bit. However, limiting bit-width of weights to 1-bit degrades the ImageNet Top-1 inference accuracy by 21.2 % in AlexNet [10]. Therefore, it is important to design hardware accelerators with an optimum bit-width to achieve an accuracy within the target range.

We simulated inference accuracy of an MNIST handwritten number in LeNet-5 [9] according to the bit-width of the weights in Table I, where pre-trained weights in the network with FP32 were quantized to integer weights. Quantization into 8-bit integer weights does not degrade the inference accuracy compared to that in the case with FP32. In addition, the accuracy degradation in the case of the network with 5-bit integer weights is negligible.

Artificial neurons in the Integrate and Fire model [13] perform a convolution by calculating the inner product of the input and weight vectors:

$$\text{Out}[x][y] = \mathbf{F}\left(b + \sum_{i=0}^{R-1}\sum_{j=0}^{S-1} \mathbf{X}[x+i][y+j] \cdot \mathbf{w}[x+i][y+j]\right)$$

$$= \mathbf{F}(\vec{\mathbf{X}} \bullet \vec{\mathbf{w}}^T + b) \quad (1)$$

where $\mathbf{X}$, $\mathbf{w}$, b, $\mathbf{F}$, and Out are an input vector, weight vector, bias, activation function, and output, respectively. If the multiplications required for the inner product are implemented via arithmetic shift and addition, a multiplier circuit that is more simple than that using the booth multiplier can be designed. Booth multiplication with an 8-bit weight produces five partial products including the addition of 1 for the 2's complement [14]. However, neural networks do not require exact algebraic calculations. Therefore, we quantized the weight into the sum of a limited number of $2^n$s, which we call the 'partial sub integer,' e.g., $7 \cdot X = 8 \cdot X - X$ with two sub-partial integers. Fig. 1 shows a representation of 8-bit integers partitioned into 2 or 3 partial sub-integers, which is expressed by

$$w \cdot X = X \cdot \sum_{i=1}^{N}(2^{n_i} \cdot w_i) \quad (w_i \in -1,0,1) \quad (2)$$

where N can be 2 or 3. It allows a multiplier to be implemented by only two- or three-barrel shift circuits. The maximum errors caused by the quantization by limiting the partial sub-integers into two and three are approximately 9 % and 2 %, respectively. Despite the calculation error in multiplication, the degradation caused by this quantization is negligible, as shown in Table I. This study employs the strategy of limiting partial sub-integers to increase throughput and reduce power consumption.

It has been found that DRAM access consumes 200 times more power than register access [15]. Inference accelerators with a spatial architecture, reported in [16]-[18], allow a computational engine to deliver the input, weight, and the Partial sum of the output (Psum) to an adjacent computational engine. Such spatial architectures reuse the input, weight, and Psum, helping to reduce DRAM access. However, computation in the Fully-Connected (FC) layer does not require input and weight reuse; thus, the spatial architecture is structurally disadvantageous for computation in the FC layer. In addition, the outputs of the layers are not reused in [16]-[18], as additional DRAM access is required for loading and storing outputs of layers.

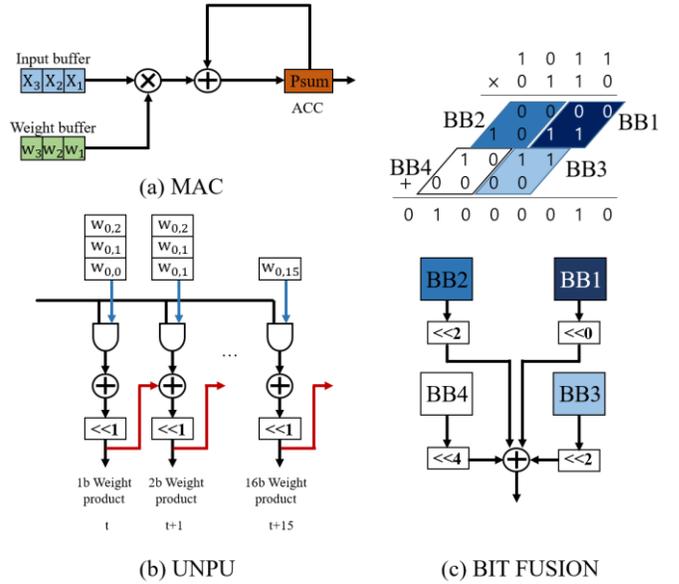

Fig. 2. Computation engine in previous works with Multiply and Accumulation (MAC) structure and shift-based multiplier. (a) MAC [19]-[21], (b) UNPU [23], and (c) BIT FUSION [24].

This paper proposes a digital neuron, a configurable inference accelerator for embedded systems. The throughput is improved by reducing the circuit area of the computational engine. The proposed digital neuron also fully utilizes computational engines with a massive parallel structure. In addition, a reuse scheme for the layer outputs is presented for reducing DRAM access.

## II. RELATED WORKS

This section reviews previous works that use multiplication with a shift circuit to reduce the complexity of a computation engine. In addition, it reviews prior arts of inference accelerators with a spatial architecture and how they reuse data and perform FC computations.

### A. Previous works with shift-based multipliers

Several hardware accelerators use the Multiply and Accumulation (MAC) structure for computing the dot product [19]-[21]. The MAC structure multiplies the input and weight with a binary multiplier (e.g. booth multiplier [14]) and accumulates the multiplied value with a binary adder (e.g. Carry Look Ahead (CLA) adder [22]), as shown in Fig. 2 (a).

A shift circuit has been adopted for arithmetic multiplication to reduce the circuit area of a multiplier [23], [24]. UNPU, a hardware accelerator reported in [23], produces 16 partial products with 16 sets of AND gate, binary adder, and 1-bit shift circuit, as shown in Fig. 2 (b). The UNPU can support multiplication with weights of all bit-widths from 1-bit to 16-bits. The BIT FUSION scheme [24] decomposes, for example, 4-bit multiplication into four 2-bit multiplications, as shown in Fig. 2 (c). By adjusting the combinations of the decomposed multiplications, this system also can support flexible bit-widths of the weights.

The proposed digital neuron for reducing the circuit area and power consumption performs 5-bit or 8-bit multiplication with 2- or 3-barrel shift circuits, respectively, as described in the Introduction. In addition, a massive parallel multi-operand adder that can reduce the circuit area and power consumption is proposed.

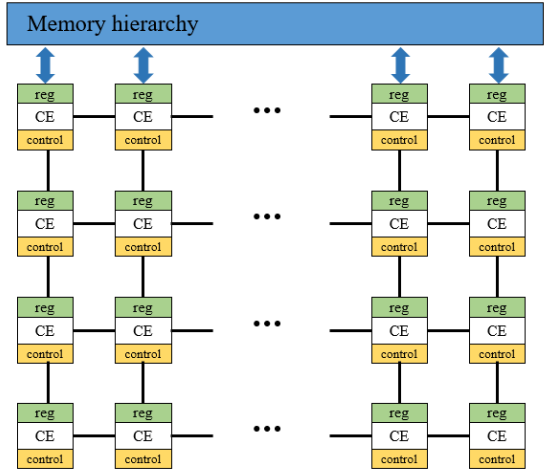

Fig. 3. Computational engine array in previous works in hardware accelerators with spatial architecture [16]-[18].

*B. Previous works on hardware accelerators with spatial architecture*

Hardware accelerators such as Eyeriss [16], DSIP [17], and COSY [18] have adopted spatial architecture. The computational engine in the spatial architecture delivers the input, weight, and Psum to an adjacent engine, as shown in Fig. 3. This data delivery between computational engines allows the reuse of the input, weight, and Psum. Therefore, in the case of computation of convolution with a filter sweep, data reuse helps to reduce DRAM access. However, in the case of computation in the FC layer, the reuse of weights is not required, as there is no sweep. Therefore, only one column of the engine array is utilized for computation in the FC layer. The massive parallel structure of the digital neuron allows full utilization of not only convolutional computation but also fully-connected computation. The studies in [16]-[18] reuse Psum, but not the output of each layer. This study presents an architecture for reusing the output of layers in order to reduce DRAM access.

### III. ARCHITECTURE OF PROPOSED DIGITAL NEURON

In this section, we present the architecture of the proposed digital neuron employing 8-bit integer weights produced by the addition of three partial sub-integers as indicated in Eq. (2). In addition, we propose a configurable system architecture that can adjust the filter size.

*A. Architecture of computational engine*

The artificial neuron of the proposed digital neuron is a computational engine that computes the dot product of two vectors digitally. This section shows an architecture of the digital neuron adopting unsigned 8-bit integer inputs and signed 8-bit integer weights with three partial sub-integers. Fig. 4 shows block diagram of the proposed digital neuron. The digital neuron performs the convolution with a 3D filter, in which $N_{ch}$

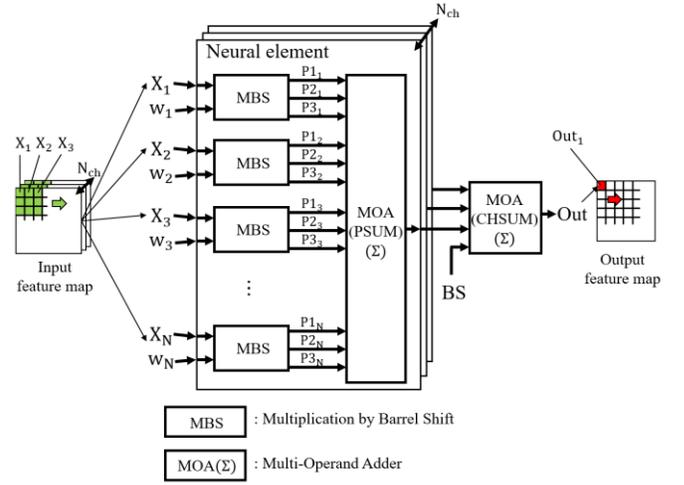

Fig. 4. Block diagram of computational engine of proposed digital neuron. The digital neuron performs convolution with a 3D filter. The digital neuron also contains M neural elements, which perform convolution with a 2D filter of one input channel. Each MBS block contains two barrel shift circuits that produce two partial products. The MOA blocks perform arithmetic sums of the input numbers.

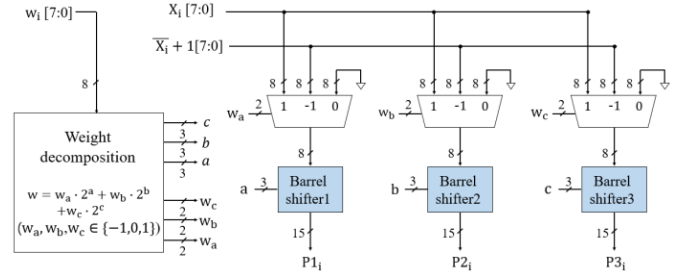

Fig. 5. Block diagram of the MBS block of proposed digital neuron. The MBS block performs multiplication. The three barrel shifters produce three partial sub-integers.

is the number of the input channels and N is the number of filtered inputs in one channel.

The Neural element block shown in Fig. 4 performs the 2D convolution of one channel in an input feature map with N inputs and N weights for each channel. The neural element consists of Multiplication by Barrel Shift (MBS) block and Multi-Operand Adder (MOA) block. The MBS block performs multiplication as expressed in Eq. (2), and the MOA block aggregates the outputs of the MBS blocks.

Each MBS block contains three barrel shift circuits that produce three partial products, i.e., P1, P2, and P3, which are $w_a 2^a$, $w_b 2^b$, and $w_c 2^c$ in Eq. (2), respectively. The MBS block receives an 8-bit weight and produces six signals, a, b, c, $w_a$, $w_b$, and $w_c$ for controlling the barrel-shift and mux circuits. The MBS block also receives a positive input, a negatized input, and 8-bit zeros. The mux circuits select one of the inputs based on the $w_i$ signal. The barrel-shift circuit multiplies the output of the mux circuit by $2^n$.

The MOA(PSUM) block collects 3N sub-partial integers (i.e. outputs of the MBS blocks), and aggregates them in massive-parallel. Fig. 6 shows a schematic of the proposed MOA(PSUM) block, where N is 25. The operational principle of the MOA block is similar that of the Wallace tree adder [25]. The MOA block groups three bits in the same column of P1, P2, and P3 vectors along the vertical direction into one group. Then, full adder arrays in each stage reduce the number of input vectors by two-thirds, and simultaneously expand their bit-width by 1. In the output of stage 10, the number of P1 and P2 vectors is reduced to two. Finally, these two vectors are added by a CLA adder. Finally, the outputs of all neural elements and a bias BS are collected and summed by the MOA(CHSUM) block.

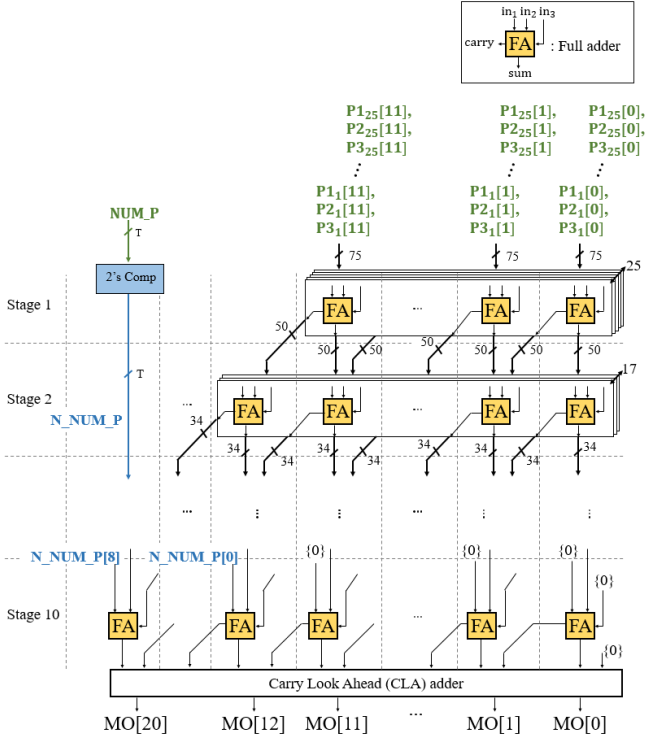

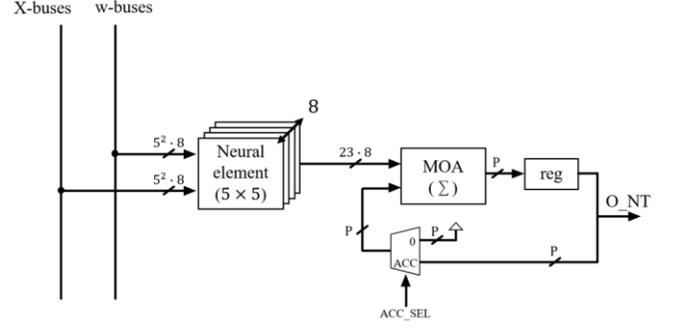

Fig. 7. Block diagram of system architecture of proposed inference accelerator.

Fig. 6. Schematic of proposed MOA(PSUM) block, where N is 25. Full adder arrays in each stage reduce the number of partial elements by two-thirds, and simultaneously expand their bit-width by 1. N_NUM_P and two's complemented-NUM_P enables the calculation of the sum of individual signed inputs. NUM_P and N_NUM_P are added in the last stage, as they are delayed than P1, and P2.

To permit the calculation of the sum of numbers with individual signs, it is necessary to perform a sign-extension of the inputs of the MOA to 18-bits, where 18 is the bit-width of the output of the MOA. However, the sign-extension increases the circuit area by ~21 %. Instead of this, the NUM_P is negatized and added to from 12-bit to 18-bit in Stage 10, where NUM_P is the number of negative partial sub-integers. This requires only a 2's complement circuit and allows the sum of numbers with individual signs to be calculated. The principle of the simplified sign-extension is described in Appendix with Fig. A1.

The proposed MOA circuit (N=25) reduces the critical path and total gates by 42 % and 36 %, respectively, compared to the summation circuit with a binary-adder-based-tree-structure with 75 inputs.

*B. System architecture of proposed inference accelerator*

The proposed inference accelerator employs a computational engine composed of 8 5×5 Neural elements, and it is allowed to accumulate O_NT according to the ACC_SEL signal, as shown in Fig. 7. The computational engine is called a Neural Tile (NT).

Fig. 8 shows the system architecture of the proposed inference accelerator. Four NTs, i.e., NT1, NT2, NT3, and NT4, perform 3D convolution in the convolutional layer and compute the dot product of two vectors in the FC layer. The NTs perform the 3D convolution and full-connection in one clock cycle.

The weights of all the layers are loaded from the DRAM to w bank registers, which are then reused in every inference. The inputs are loaded from the DRAM to the Input feature map registers for one inference. The four outputs of the NTs, which are O_NT1, O_NT2, O_NT3, and O_NT4, are added by two CLA adders. The outputs of the two CLA adders, O_NT1P2 and O_NT3P4, are added by another CLA adder to obtain the O_NTSUM. The utilization of these signals is described in detail in sub-section C. The calculation results of the computational engines are stored in Output feature map registers after activation in the Activation (ReLU) block. A pooling operation is performed on the outputs of the Output feature map register. The pooled outputs are stored in the Input feature map.

The outputs of the Input feature map are assigned to the inputs of the NT1 to NT4 blocks. Fig. 9 shows the data transfer from the Input feature map to the X-bus and the behavior of w bank. Elements of the filtered area that require dot product computation is assigned to X-bus from the Input feature map. After one clock cycle, the filtered area in the input feature map must be shifted one column to the right. This is implemented by updating one column in the X-bus, i.e., the leftmost column from the previous clock cycle in the Input feature map is deleted and the rightmost column of the next clock cycle in loaded in its place. This is done by increasing the address (ADDR) of the column of the Input feature map by 5. Columns that are not updated do not consume dynamic power. Therefore, this helps to reduces the power consumption. The w bank rotates to the right column-by-column according to the sweep of the filter. This is implemented by connecting shift registers in the horizontal direction from the rightmost columns to the leftmost columns. At the beginning of the next row of the filter sweep, the whole filtered area of the Input feature map is assigned to the X-bus, and the w-bus is updated with the initial columns of CLK1.

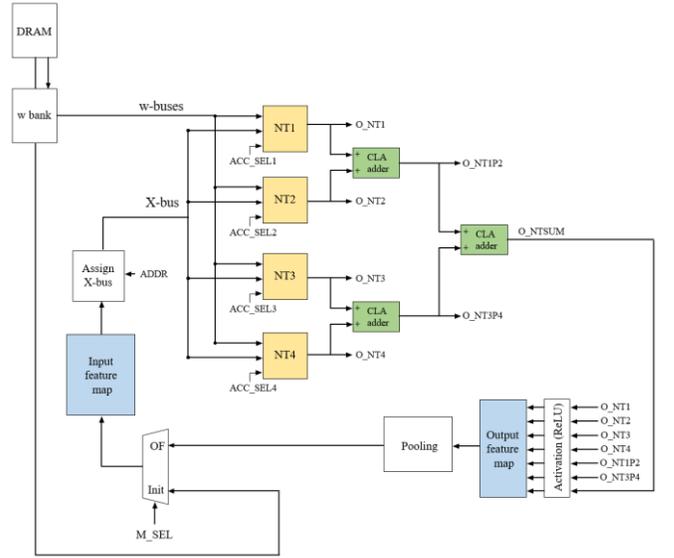

Fig. 8. Block diagram of system architecture of proposed inference accelerator.

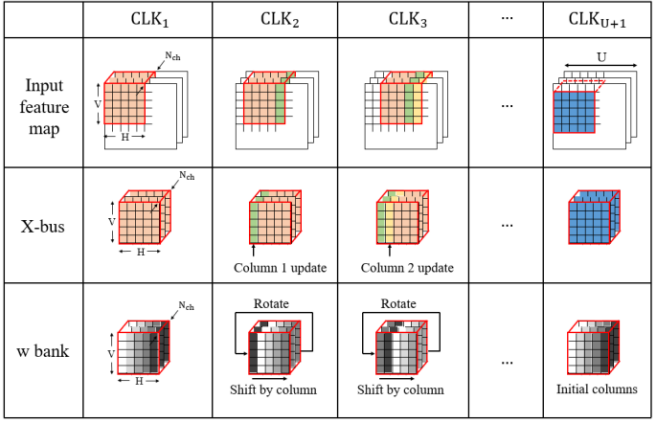

Fig. 9. Data transfer from Input feature map to X-bus, and column rotation of w bank. The Assign X-bus block assigns elements to the X-bus by controlling the address signal (ADDR).

*C. Configurable scheme for adjusting filter size*

Fig. 10 shows an example of different applicable filter sizes in the proposed inference accelerator depending on the input area of the NTs and the combination of adders. Case 1 shows the example of the application of a $5 \times 5 \times 32$ filter, in which each of the four NTs receives a quarter of the filter divided based on the depth and the dot product of the quarter is calculated. Then, the outputs of the four NTs are summed by three CLA adders. In this case, the Output feature map collects four O_NTSUMs and delivers them to the Pooling block.

Case 2 shows an example of the application of a $7 \times 7 \times 16$ filter, where NT1 and NT2 receive a half and the remainder of the $7 \times 7$ filter, i.e., $[0:24] \times [0:7]$ and $[0:48] \times [0:7]$. Likewise, NT3 and NT4 also receive a half and the remainder of the $7 \times 7$ filter, i.e., $[0:24] \times [8:15]$ and $[0:48] \times [8:15]$. The four outputs of the four NTs are summed by three CLA adders. In this case also, the output feature map collects four O_NTSUMs and delivers them to the Pooling block.

Case 3 shows an example of the application of a $9 \times 9 \times 8$ filter, in which NT1, NT2, NT3, and NT4 receive four fractions

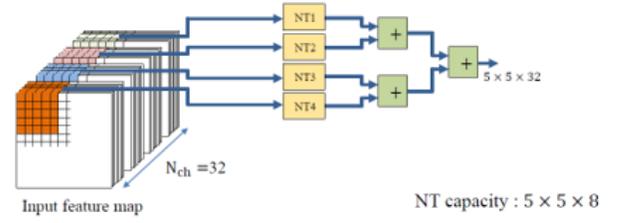

(a) Case 1 - 5×5×32

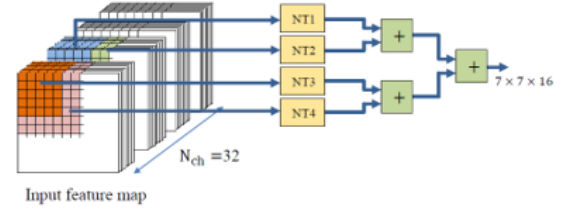

(b) Case 2 - 7×7×16

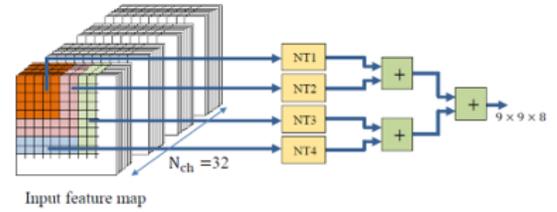

(c) Case 3 - 9×9×8

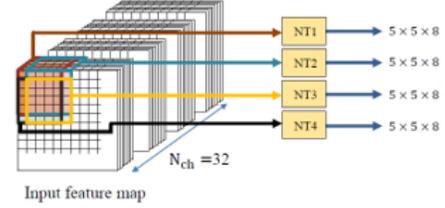

(d) Case 4 – Four 5×5×8

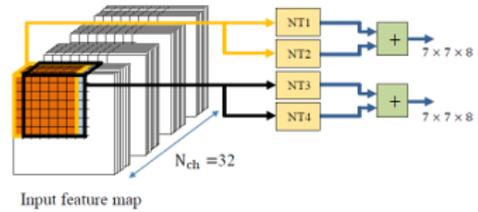

(e) Case 5 – Tow 7×7×8

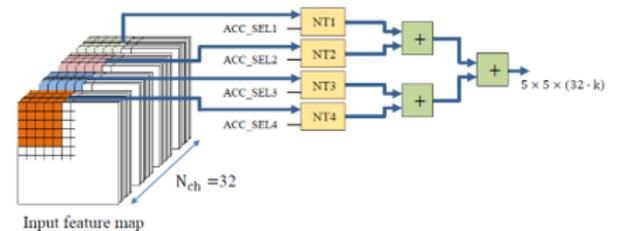

(f) Case 6 –5×5×128

Fig. 10. Example of applicable different filter sizes in the proposed inference accelerator.

of the filter, i.e., $[0:24] \times [0:7]$, $[25:49] \times [0:7]$, $[50:74] \times [0:7]$, and $[75:80] \times [0:7]$. In this case also, the output feature map collects four O_NTSUMs and delivers them to the Pooling block.

Case 4 shows an example the application of a $5 \times 5 \times 8$

TABLE II
Performance of the implementation of the proposed inference accelerator in FPGA DE2-115 board for MNIST handwritten number recognition operating LeNet-5 [6].

| Performance of the proposed system architecture | |
|---|---|
| Inference accuracy | 98.92 % |
| Number of clocks required in computing dot product of NTs (25 MHz) | 1 clock |
| Number of clocks/computation time (25 MHz) for inference of one image | 2,384 clocks/ 95 μs |
| Frame rate (25 MHz) | 19 k frame/s |

TABLE III
Comparison of performance with an artificial neuron in a prior art computing dot product of two vectors with 50 components.

| | Adder tree with booth multiplier (N=25) [4] | YodaNN (N=25) [16] | Proposed neural element (N=25) | |
|---|---|---|---|---|
| Weight precision | 5-bit (exact) | 1-bit | 5-bit integers (2 partial sub-integers) | 8-bit integers (3 partial sub-integers) |
| Activation precision | 8-bit | 8-bit | 8-bit | 8-bit |
| Critical path | 57 | 40 | 41 | 46 |
| Total gates | 15,688 | 4,715 | 8,201 | 12,502 |
| Simulated power consumption | 4.940 mW | 1.452 mW | 2.576 mW | 3.953 mW |

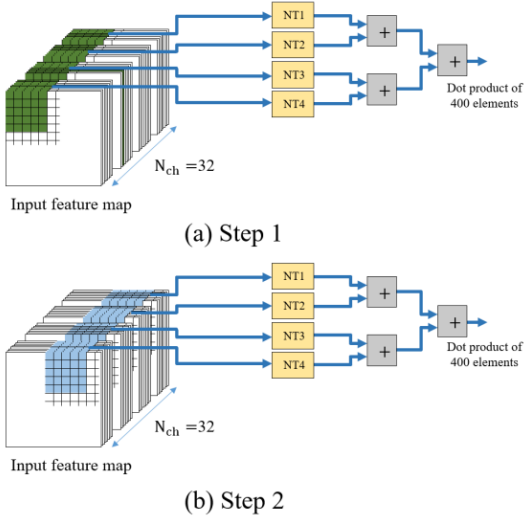

Fig. 11. Computation steps in fully connected layer: (a) Step1 and (b) Step2.

filter, where each of the four NTs receives a separate $5 \times 5 \times 8$ area in the Input feature map. The four NTs perform four $5 \times 5 \times 8$ convolutions in parallel, and the four outputs of the NTs, O_NT1, O_NT2, O_NT3, and O_NT4 are stored in the Output feature map. The Output feature map collects the four O_NTs and delivers them to the Pooling block.

Case 5 shows an example of the application of a $7 \times 7 \times 8$ filter, where two groups of two NTs receive two separate $7 \times 7 \times 8$ areas in the Input feature map. The outputs of NT1 and NT2 are added by one CLA adder and the outputs of NT3 and NT4 are added by another CLA adder. These two outputs of the CLA adders, O_NT1P2 and O_NT3P4 are stored in the Output feature map. The Output feature map collects four O_NT1P2/O_NT3P4s and delivers them to the Pooling block.

Case 6 shows an example of the application of a $5 \times 5 \times 128$ filter, where the depth of the filter exceeds the summed depth of four NTs. In this case, the ACC_SEL signal allows the outputs of the NTs to be accumulated. The depth of the filter is 128. Therefore, the outputs are accumulated four times in total. However, in this case, all the signals of the X-bus and w-bus should be updated. Thus, it permits calculation, but increases power consumption.

### D. Computation in fully connected layer

Fig. 11 shows the computation steps of the FC layer. The 1D vectors of the FC layer are stored as $5 \times 5 \times 32$ elements of the Input feature map. In step 1, the digital neuron calculates the dot products of 400 elements, and in step 2, it calculates the dot products of the next 400 elements. The proposed massive parallel architecture does not degrade the throughput in the computation in the FC layer.

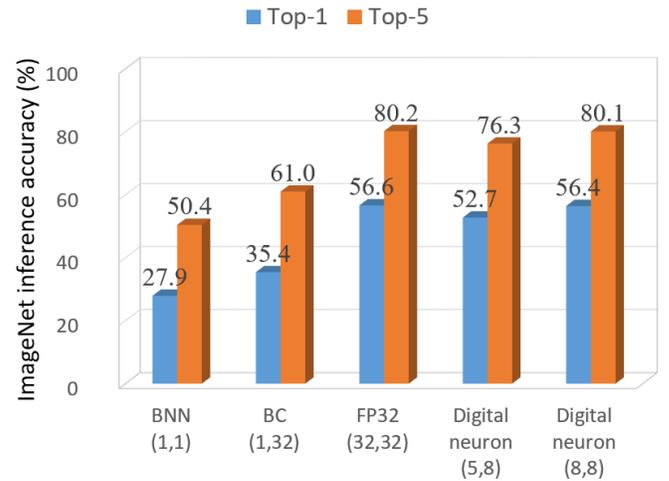

Fig. 12. Comparison of ImageNet inference accuracy of the proposed digital neurons with those in previous works, i.e., FP32, Binarized Neural Network (BNN) [26], and BinaryConnect (BC) [12]. In the proposed digital neurons, 5-bit and 8-bit weights comprises two and three partial sub-integers, respectively.

## IV. EXPERIMENTAL RESULTS

In this section, we implement a CNN network, Lenet-5 [9], for MNIST handwritten number recognition using the proposed digital neuron on the FPGA DE2-115 board and explain the performance of the implemented system.

Weights were trained with LeNet-5 with FP32 precision. To integerize the pre-trained weights, all the weights were multiplied by 16 and quantized into 5-bit integer weights with two partial sub-integers. The quantized integer weights are delivered to the DRAM on the FPGA board. As the weights were multiplied by 16 to integerize them, the least significant 4-bits of the output of the digital neuron were truncated to normalize them.

Table II summarizes the performance of the implemented hardware inference accelerator. An inference accuracy of 98.95 % is achieved when training the MNIST database. Only one clock at 25 MHz is required for computing the dot product of the NTs. The number of clocks and computation time required for the inference of one image is 2,384 clocks at 25 MHz and 95 μs, respectively. The frame rate for inference is 19 k frame/s.

TABLE IV
Performance comparison of inference accelerators with those in previous works

|  | Eyeriss [16] | ConvNet [27] | DSIP [17] | This work (Simulation) |
|---|---|---|---|---|
| Weight precision | 16-bit fixed point | 16-bit fixed point | 16-bit fixed point | INT5 |
| Activation precision | 16-bit fixed point | 16-bit fixed point | 16-bit fixed point | INT8 |
| Number of MACs | 168 | 256 | 64 | 800 |
| Technology | 65 nm | 40 nm | 65 nm | 65 nm |
| Operating voltage | 1.0 V | 0.9 V | 1.2 V | 1.2 V |
| power consumption | 278 mW | 274 mW | 88.6 mW | 212 mW |
| Frequency | 250 MHz | 204 MHz | 250 MHz | 200 MHz |
| Throughput | 23.1 GMACS | 52.2 GMACS | 30.1 GMACS | 160 GMACS |
| GMACS/W | 83.1 GMACS/W | 190.6 GMACS/W | 136.8 GMACS/W | 754.7 GMACS/W |

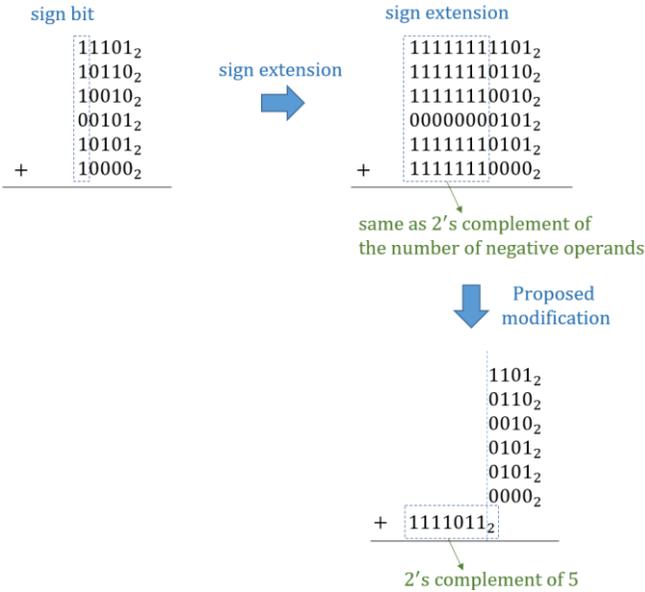

Fig. A1. Example of summation of six 5-bit binary numbers. This example describes summation of extended sign bits is same as 2's complement of the number of negative operands. Therefore, sign extension can be replaced to addition of one binary number.

## V. DISCUSSION

In this section, we compare performance of the proposed computational engine and inference accuracy with those of prior works. We also compare the performance of the proposed inference accelerator with those in prior works.

To compare the performance, three kinds of dot product-computing artificial neurons, i.e., adder tree with booth multiplier, YodaNN (1-bit weight), and the proposed neural element were simulated. Table III summarizes the results of the comparison. The proposed neural element surpasses the adder tree with booth multiplier in terms of the gate delay and power consumption by ~1.4x and ~1.9x, respectively. In addition, the circuit area is reduced based on a comparison of the total gates of the digital neuron and the adder tree scheme. However, the proposed neural element consumes ~1.8x more power than YodaNN, and the bit-width of the proposed digital neuron is 5x that of YodaNN.

The quantization of the weights into binary bit results in a significant reduction in the computational power and storage requirements. However, excessive reduction in the bit-width of the weights results in the degradation of the inference accuracy. Fig. 12 shows a comparison of the ImageNet inference accuracy of the proposed digital neuron and previous works, i.e., FP32, Binarized Neural Networks (BNN) [26], and BC [12]. As shown in the figure, the quantization of the weights into 1-bit results in a decrease in the inference accuracy by more than 20 % compared to that of FP32. However, for AlexNet with the proposed digital neuron (5,8), there is an additional error of 3.9%/3.9% in the Top-1/Top-5 accuracy, compared to that in the FP32 case. The additional error can be reduced by increasing the precision. For AlexNet with the proposed digital neuron (8,8), there is an additional error of only 0.2%/0.1% in the Top-1/Top-5 accuracy.

Table IV shows a comparison of the performance of the simulated inference accelerators with those in previous works, where the simulation was performed using HSPICE with 65 nm CMOS process technology at an operation voltage of 1.2 V and frequency of 200 MHz. The digital neuron had a power consumption of 212 mW while operating convolutional layer 2 of AlexNet. The throughput of the digital neuron was 160 Giga MACs (GMACS) with 754.4 GMACS/W. These values surpass those of the inference accelerators in previous studies. For example, compared with that of ConvNet [27], the throughput per watt for the digital neuron in this work is ~3.96x more.

## VI. CONCLUSION

We proposed and implemented a digital neuron, a configurable hardware inference accelerator with integer weights and inputs for embedded systems. A quantization scheme that partitions the integer weight into a limited number of partial sub-integers was applied. The degradation in the ImageNet inference accuracy is negligible with the weight partitioned into 3 partial sub-integers, i.e., ~0.2%. The proposed massive parallel architecture does not reduce the throughput in the computation in the FC layer. With the proposed quantization scheme and massive parallel summation technique, the power consumption of the computational engine was reduced by ~1.9x compared to that of the conventional engine. The operation was verified by implementing the digital neuron in the FPGA DE2-115 board. The digital neuron was simulated at an operating frequency of 200 MHz to compare the performance in this work with those of prior arts. The throughput per watt of the simulated digital neuron is ~3.96x more than in previous work.

APPENDIX

The Appendix describes the principle of the proposed sign extension of the MOA circuit. Fig. A1 shows an example of the summation of six 5-bit binary numbers. In the conventional method, negative numbers should extend 1s and positive numbers should extend 0s. However, the binary numbers of the extended 1s and 0s are the same as -1 and 0, respectively. Therefore, the summation of the extended bits can be replaced with the 2's complement of the number of negative numbers.